\documentclass[12pt,a4paper]{article}
\usepackage[cp1251]{inputenc}
\usepackage[english]{babel}
\usepackage{graphicx}
\usepackage{amsmath}
\date{} \lccode`\-=`\-
\title{Relation between full traces of Green functions
for  initial and  Darboux transformed  Dirac problems}
\author{Ekaterina Pozdeeva\\
 {\it Department of Quantum Field Theory,}\\
 {\it Tomsk State University, 36 Lenin Avenue}\\
{\it Tomsk, 634050, Russia}\\
{\it ekatpozdeeva@mail.ru} }
\begin{document}
\selectlanguage{english} \maketitle

\begin{abstract}
We establish the relation between full traces of the Green functions
for some initial and the Darboux transformed one-dimensional two
component Dirac problems with the most general form of potential.
The result is used to check the completeness of set of wave functions
obtained by the Darboux transformation of the eigenfunctions  set for
the initial Dirac problem with some typical boundary conditions.\\

Keywords: Dirac equation; Green function; Darboux transformation.
\end{abstract}
\section{Introduction}

Supersymmetric quantum mechanics (SQM)  \cite{Witten,2} provides an interesting framework within which  to
analyse quantum problems. In particular, it allows one to investigate the spectral properties
 of a wide class of quantum  models and to generate new systems with given spectra.
  SQM gives new insight into the problem of spectral equivalence of Hamiltonians, which,
  historically, has been constructed as a factorization method in quantum mechanics \cite{3}
 and as Darboux-Crum transformations in mathematical physics \cite{4}.

 The Darboux transformation
method for the one-di\-me\-n\-sio\-nal sta\-tio\-na\-ry Dirac
equation is equivalent to the underlying quadratic supersymmetry and
the factorization properties of the Dirac equation
\cite{annphys2003v305p151}.
Application of this method  to the Dirac
equation is studied in the papers \cite{Eurjphys,preprint}.

In the previous paper \cite{preprint} we have established the connection between
Green functions of initial and Darboux transformed one-dimensional
two component Dirac equations for the case of  especial matrix
structure of the potential. Here we consider the same
problem for the arbitrary matrix structure of the interaction
Hamiltonian.

\section{Green function of initial problem}

We consider the Dirac equation
\begin{equation}
\label{1} h_0(x)\psi(x)=E\psi(x), \qquad \psi=(\psi_1,\psi_2)^T
\end{equation}
with the  Hamiltonian of the form
\begin{equation}\label{2}h_0=i\sigma_2\partial_x+V(x),\end{equation}
 \begin{equation}\label{3}
    V_0(x)=\omega(x)I+(m+S(x))\sigma_3+q(x)\sigma_1,
\end{equation}
where $\omega(x)$, $S(x)$ and $q(x)$ are real functions of $x$; $m$
is the mass of a particle; $\sigma_1$, $\sigma_2,$ $\sigma_3$ are usual Pauli
matrices. This equation have two linearly independent solutions.

Denote them
by $\psi$ and $\varphi.$ Introduce the Wronskian of this solutions
with the help of equation
\begin{eqnarray}
W\{\psi(x),\varphi(x)\}&=&\psi_1(x)\varphi_2(x)-\psi_2(x)\varphi_1(x).
\end{eqnarray}
Let us prove that it doesn't depend on $x$. For this aim represent
$W$ in the form:
\begin{eqnarray}
W&=&\psi^T\gamma\varphi, \quad \gamma=i\sigma_2=\left(
                                                  \begin{array}{cc}
                                                    0 & 1 \\
                                                    -1 & 0 \\
                                                  \end{array}
                                                \right), \quad
                                                \gamma^T=-\gamma.
\end{eqnarray}
Then \begin{eqnarray}
     W'&=&(\psi^T)'\gamma\varphi+\psi^T\gamma\varphi'=\psi^T\gamma\varphi'-(\gamma\psi')^T\varphi.
     \end{eqnarray}
Taking into account the Dirac equation
\begin{eqnarray}
\gamma\varphi'&=&E\varphi-V\varphi, \\
 \gamma\psi' &=& E\psi-V\psi,
\end{eqnarray}
we have \begin{eqnarray}
     W'&=&\psi^T(E\varphi-V\varphi)-(E\psi^T-(V\psi)^T)\varphi=\psi^T(E-V-E+V^T)\varphi.
     \end{eqnarray}

Since $V^T=V$, $W'=0$.

 Now we are in position to construct Green
function of problem under consideration. It is the solution of
inhomogeneous equation
\begin{eqnarray}
(H-E)G(x,y)&=&\delta(x-y).
\end{eqnarray}
It is easy to check that
\begin{eqnarray} \label{G}
G(x,y)=\frac{\varphi(x)\psi^T(y)\Theta(x-y)+\psi(x)\varphi^T(y)\Theta(y-x)}{W}.
\end{eqnarray}
Besides this for $G(x,y)$ the spectral representation
\begin{eqnarray}
G(x,y)=\Sigma_n\frac{\varphi_n(x)\varphi_n^T(y)}{E-E_n}
\end{eqnarray}
is valid.  Here $\varphi_n(x)$ is complete orthonormal system of
eigenfunctions of the $h_0$ with eigenvalues $E_n$.
\section{Darboux transformation of the Green function}
Let us to construct the $2\times2$  matrix $u$,  consists of two two-component
eigenfunctions $\varphi_{n_1}$, $\varphi_{n_2}$ of the Dirac Hamiltonian $h_0$ corresponding eigenvalues
$E_{n_1}$, $E_{n_2}$:
\begin{eqnarray}
\label{u}u&=&(\varphi_{n_1}, \varphi_{n_2}).
\end{eqnarray}
Thus, the function $u$ is a solution of the Dirac
equation \begin{eqnarray}
         h_0u&=&u\Lambda,\qquad \Lambda=diag(E_{n_1},E_{n_2}).
         \end{eqnarray}
The operator \begin{eqnarray}
L&=&\frac{d}{dx}-u_xu^{-1},\qquad u_x=\frac{du}{dx}
\end{eqnarray}
allows us to generate the solutions $\tilde{\varphi}_n$
of the transformed Dirac equation
\begin{eqnarray}
h_1\tilde{\varphi}_n&=&E_n\tilde{\varphi}_n,
\end{eqnarray}
from solutions of the initial Dirac equation.
The Hamiltonian  and corresponding solutions of the transformed Dirac equation are the following
\begin{eqnarray}
      \tilde{\varphi}_n&=&L\varphi_n \\
        h_1&=&\gamma\frac{d}{dx}+V_1,\qquad V_1=V_0+[\gamma,u_xu^{-1}],
      \end{eqnarray}
where $\varphi_n$ are solutions of the initial equation.

It is easy to check that $L\varphi_{n_1}=0$, $L\varphi_{n_2}=0.$
The spectrum of $h_1$ doesn't contain eigenvalue $E_{n_1}$, $E_{n_2}.$

The question arises: are the eigenfunctions $\tilde{\psi}_n$ $(n\neq
1,2)$ form the full system.
In order to answer to this question it is necessary to construct Green
function $G_1$
\begin{eqnarray}
(h_1-E)G_1&=&\delta(x-y)
\end{eqnarray}
and to calculate the expression
\begin{eqnarray}
\label{A}
  A&=&tr\int(G_1(x,x)-G_0(x,x))dx+\frac{1}{E_{n_1}-E}+\frac{1}{E_{n_2}-E}.
\end{eqnarray}

If $A=0$, the set of $\tilde{\psi}$ form the complete system.

 Early expression, similar to \eqref{A},
was considered for  Schr\"o\-din\-ger equation in \cite{Sukumar}.

Evidently that
\begin{eqnarray}
G_1(x,y)&=&\frac{\tilde{\varphi}(x)\tilde{\psi}^T(y)\Theta(x-y)+\tilde{\psi}(x)\tilde{\varphi}^T(y)\Theta(y-x)}{\widetilde{W}},\\
  \widetilde{W}&=&\tilde{\psi}^T(x)\gamma\tilde{\varphi}(x).
\end{eqnarray}

It can be proved (see Appendix A) that $\tilde{W}=(E-E_{n_1})(E-E_{n_2})W.$

Consider \begin{eqnarray}
       tr\{\tilde{\psi}(x)\tilde{\varphi}^T(x)\}=\tilde{\psi}^T(x)\tilde{\varphi}(x)=(L\psi(x))^TL\varphi(x).
         \end{eqnarray}
It is evidently that
\begin{eqnarray}
  \tilde{\psi}&=&L\psi=\psi'(x)-u_xu^{-1}\psi(x)\equiv u(x)(u^{-1}\psi)'
\end{eqnarray}
and $$\tilde{\psi}^T(x)=[(u^{-1}\psi)']^Tu^T,$$
\begin{eqnarray}
\tilde{\psi}^T(x)\tilde{\varphi}(x)&=&(u^{-1}\psi)^T)'u^T\tilde{\varphi}=\{\psi^T(u^{-1})^Tu^T\tilde{\varphi}\}'-\{\psi^T(u^{-1})^T(u^T\tilde{\varphi})'\}.
\end{eqnarray}
Since $u$ is a real matrix (this follows from reality of all coefficients of the Dirac equation
rewritten in component form) $u^T=u^+.$ The last form in the equation can be represent in the form
\begin{eqnarray}
\psi^TL^+\tilde{\varphi}, \qquad L^+=-\partial_x-(u_xu^{-1})^+.
\end{eqnarray}
Taking into consideration that $\tilde{\varphi}=L\varphi,$ factorization relation \cite{annphys2003v305p151}
$$L^+L=(h_0-E_{n_1})(h_0-E_{n_2}),$$
and $$h_0\varphi=E\varphi,$$ we have
\begin{eqnarray}
\tilde{\psi}^T\tilde{\varphi}&=&\psi^T\tilde{\varphi}+(E-E_{n_1})(E-E_{n_2})\psi^T\varphi.
\end{eqnarray}
Thus, \begin{eqnarray}
trG_1(x,x)=\frac{\tilde{\psi}^T\tilde{\varphi}}{\widetilde{W}}&=&\frac{(\psi^T\tilde{\varphi})'+(E-E_{n_1})(E-E_{n_2})(\psi^T\varphi)}{(E-E_{n_1})(E-E_{n_2})W}\nonumber\\
&&=\frac{(\psi^T\tilde{\varphi})'}{\widetilde{W}}+tr\{G(x,x)\}.
\end{eqnarray}
Then the trace of difference of the Green functions is as follows
\begin{eqnarray}
\label{G1G0}
tr[G_1(x,x)-G_0(x,x)]&=&\frac{(\psi^T\tilde{\varphi})'}{\widetilde{W}}.
\end{eqnarray}
Evidently that \eqref{G1G0} can also be presented in the form
\begin{eqnarray}
tr[G_1(x,x)-G_0(x,x)]&=&\frac{(\tilde{\psi}^T\varphi)'}{\widetilde{W}}.
\end{eqnarray}
Thus, \begin{eqnarray}
    tr\{\psi^T\tilde{\varphi}-\tilde{\psi}^T\varphi\}&=&C,\qquad C'=0.
     \end{eqnarray}
Let us prove that
\begin{eqnarray}
\label{S}
C=(2E-E_{n_1}-E_{n_2})W.
\end{eqnarray}
Taking into account the Dirac equations for $\psi$, $\varphi$ and $u$
\begin{eqnarray}
\gamma\psi'=(E-V), \qquad \gamma\varphi'=(E-V)\varphi,\qquad \gamma u_xu^{-1}=\Omega-V,
\end{eqnarray}
where $\Omega=u\Lambda u^{-1}$ it is easy to see that
\begin{eqnarray}
\tilde{\psi}^T\varphi-\psi^T\tilde{\varphi}&=&\psi^T[(\Omega^T-E)\gamma^T-\gamma(\Omega-E)]\nonumber\\
&=&2E\psi^T\gamma\varphi-\psi^T(\Omega^T\gamma+\gamma\Omega)\varphi.
\end{eqnarray}
Since $$\Omega^T\gamma+\gamma\Omega=\gamma tr\Omega$$
and
$$tr\Omega=E_{n_1}+E_{n_2}$$ the relation $$\psi^T\tilde{\varphi}-\tilde{\psi}^T\varphi=(2E-E_{n_1}-E_{n_2})W$$
is proved.

Taking into account this relation one can obtain
\begin{eqnarray}
\label{37}
tr\int_a^b[G_1(x,x)-G_0(x,x)]dx&=&\frac{1}{E-E_{n_1}}+\frac{1}{E-E_{n_2}}\nonumber\\
&+&\tilde{\psi}^T(b)\varphi(b)-\psi^T(a)\tilde{\varphi}(a)
\end{eqnarray}
or
\begin{eqnarray}
\label{38}
A=tr\int_a^b[G_1(x,x)-G_0(x,x)]dx&+&\frac{1}{E_{n_1}-E}+\frac{1}{E_{n_2}-E}\nonumber\\
&=&\tilde{\psi}^T(b)\varphi(b)-\psi^T(a)\tilde{\varphi}(a).\end{eqnarray}
If  right side of this relation equal to zero the wave functions $\tilde{\varphi}_n$
form complete system of eigenfunctions of $h_1$. In opposite case
this system is incomplete.
\section{Examples}
In this Section we consider some examples of the application of \eqref{38} to the
 Dirac problem with homogeneous boundary conditions on the finite interval
 $x\in[a,b].$ More explicitly, we look for the solutions of the Dirac equation
 \begin{eqnarray}
 h\psi&=&E\psi,
 \end{eqnarray}
that satisfy  the boundary conditions of type
\begin{eqnarray}
\psi_1(a)\cos{\alpha}+\psi_2(a)\sin{\alpha}&=&0, \\
\varphi_1(b)\cos{\beta}+\varphi_2(b)\sin{\beta}&=&0.
\end{eqnarray}

We start with the consideration of the case of free initial Hamiltonian
\begin{eqnarray}
h_0&=&i\sigma_2\frac{d}{dx}+m\sigma_3
\end{eqnarray}
and consider four types of boundary conditions:
(i) $\alpha=\beta=0$; (ii) $\alpha=\beta=\pi/2$; (iii) $\alpha=0,$ $\beta=\pi/2$; (iv) $\alpha=\pi/2$, $\beta=0.$

Without the loss of generality we can put $a=0$, $b=-1.$\\

{\bf Case (i)}. The  eigenfunction of $h_0$ consists of two branches: (a)
 positive eigenvalues and (b) negative eigenvalues. Positive eigenvalues have a form:
\begin{eqnarray}
E_n&=&\sqrt{k_n^2+m^2}, \qquad k_n=n\pi, \qquad n=1,2,...
\end{eqnarray}
The corresponding eigenfunctions are
\begin{eqnarray}
\varphi_1^{(n)}&=&\sin{(k_nx)},\qquad \varphi_2^{(n)}=-\frac{-k_n\cos{k_nx}}{E_n+m}.
\end{eqnarray}
Negative eigenvalues are as follows:
\begin{eqnarray}
\bar{E}&=&-\sqrt{k_n^2+m^2}, \qquad k_n=n\pi,\qquad n=0,1,2,...
\end{eqnarray}
Corresponding eigenfunctions are
\begin{eqnarray}
\bar{\psi}_1^{(n)}&=&\sin{(k_nx)} ,\qquad \psi_2^{(n)}=-\frac{k_n\cos{(k_nx)}}{E+m}, \qquad n=1,2,... \\
\end{eqnarray}
$$ \bar{\psi}_1^{(0)}=0,\qquad  \psi_2^{(0)}=C,$$
where $C$ is an  arbitrary constant.

The independent solution of the equation
\begin{eqnarray}
h_0\psi&=&E\psi,\qquad E\not=(E_n,\bar{E}_n)
\end{eqnarray}
are \begin{eqnarray}
    \psi_1&=&\sin{(kx)},\qquad \psi_2=-\frac{k\cos(kx)}{E+m}, \qquad k=\sqrt{E^2-m^2}, \\
      \varphi_1&=&\sin{(kx-k)}, \qquad \varphi_2=-\frac{k\cos(kx-x)}{E+m}.\end{eqnarray}
Then the full trace of the Green function is
\begin{eqnarray}
tr\int G_0(x,x)dx&=&\frac{E}{k}\cot(k)-\frac{m}{k^2}.
\end{eqnarray}
Taking into account the following relation (see Appendix B)
\begin{eqnarray}
\label{1fromB}
\cot(k)&=&\frac{1}{k}+\sum_{n=1}^\infty\frac{2k}{k^2-n^2\pi^2},
\end{eqnarray}
we get
\begin{eqnarray}
\frac{E}{k}\cot(k)-\frac{m}{k^2}&=&2E\sum_{n=1}^\infty\frac{1}{k^2-n^2\pi^2}+\frac{E-m}{k^2}\nonumber\\
&=&2E\sum_{n=1}^\infty\frac{1}{E^2-E_n^2}+\frac{1}{E+m}\nonumber\\
&=&\sum_{n=1}^\infty\left(\frac{1}{E-E_n}+\frac{1}{E-E_n}\right)+\frac{1}{E+m},\nonumber
\end{eqnarray}
that is just the spectral representation of the $tr\int_0^1G_0(x,x)dx.$

Now we consider the Darboux transformation. The solutions of transformed equation are following
\begin{eqnarray}
\tilde{\psi}&=&\psi'-u_xu^{-1}\psi,
\end{eqnarray}
where $u$ is the transformation matrix. For the construction of $u$ we choose the pair of the eigenfunctions
 $\bar{\psi}^{(0)}$ and ${\psi}^{(1)}$ or pair ${\psi}^{(0)}$ and $\bar{\psi}^{(1)}.$

 In the first case the transformation matrix reads
 \begin{eqnarray}
 u&=&\left(
         \begin{array}{cc}
           0 & \sin(k_1x) \\
           C & -\frac{k_1\cos(k_1x)}{E_1+m} \\
         \end{array}
       \right),\quad
   u_xu^{-1}=\left(
                   \begin{array}{cc}
                     k_1\cot{k_1x} & 0 \\
                     (E-E_1) & 0 \\
                   \end{array}
                 \right).
 \end{eqnarray}

In the second case it reads
 \begin{eqnarray}
 u&=&\left(
         \begin{array}{cc}
           0 & \sin(k_1x) \\
           C & -\frac{k_1\cos(k_1x)}{E_1+m} \\
         \end{array}
       \right),\quad
   u_xu^{-1}=\left(
                   \begin{array}{cc}
                     k_1\cot{k_1x} & 0 \\
                     (\bar{E}_1-m) & 0 \\
                   \end{array}
                 \right).
 \end{eqnarray}

This leads to the following expressions for components of  functions
 $$\tilde{\psi}(x)=L\psi,\quad and \quad\tilde{\varphi}(x)=L\varphi:$$
\begin{eqnarray}
\tilde{\psi}_1&=&k\cos{(kx)}-\pi\cot(\pi x)\sin(kx), \\
\tilde{\psi}_2&=&(E-E_1)\sin(kx)\quad (or~\tilde{\psi}_2=(E-\bar{E}_1))\sin(kx),\\
\tilde{\varphi}_1&=&k\cos{(kx-k)}-\pi\cot(\pi x)\sin(kx-k), \\
\tilde{\varphi}_2&=&(E-E_1)\sin(kx-k) \quad (or~\tilde{\psi}_2=(E-\bar{E}_1))\sin(kx-k).
\end{eqnarray}

 It is interesting to note that at the left side ($x=0$) of the interval  both components
  of $\tilde{\psi}$ ($\tilde{\psi}_1$ and  $\tilde{\psi}_2$) are zero, that is $\tilde{\psi}(a)=0.$
  Similarly, $\tilde{\varphi}_1(1)=\tilde{\varphi}_2(1)=0$ or $\tilde{\varphi}(b)=0.$
 Thus, the right side of eq. \eqref{38} is zero and consequently system of function $L\phi^{(n)}$, $L\bar{\phi}^{(n)}$
  is complete.\\

 {\bf Case (ii)}. The eigenspectrum again  consists of two branches:
\begin{itemize}
 \item[(a)] $E_n=\sqrt{k_n^2+m^2},\quad k_n=n\pi, \quad n=0,1,...$
 \item[(b)] $\bar{E}_n=-\sqrt{k_n^2+m^2},\quad k_n=n\pi, \quad n=0,1,...$
\end{itemize}
Solutions $\psi$, $\varphi$ reads:
\begin{eqnarray}
\psi_2(x)&=&\sin(kx), \quad \psi_1(x)=\frac{k\cos(kx)}{E-m},\quad k=\sqrt{E^2-m^2} \\
  \varphi_2&=&\sin(kx-k), \quad \varphi_1(x)=\frac{k\cos(kx-k)}{E-m}.
\end{eqnarray}
Then  the corresponding full trace of the Green function may be represent in the form:\begin{eqnarray}
tr\int G(x,x)dx&=&\frac{E}{k}\cot(k)+\frac{m}{k^2}=\sum_{n=1}^\infty \left(\frac{1}{E-E_n}+\frac{1}{E+E_n}\right)+\frac{1}{E-m}.
\end{eqnarray}
The transformation matrix $u$ is
\begin{eqnarray}
u&=&\left(
      \begin{array}{cc}
        1 & k_1\cos(k_1x)/(E_1-m) \\
        0 & \sin(k_1x) \\
      \end{array}
    \right)~or~u=\left(
                              \begin{array}{cc}
                                1 & k_1\cos(k_1x)/(E_1-m) \\
                                0 & \sin(k_1x) \\
                              \end{array}
                            \right)
\end{eqnarray}
and components of solutions of the transformed Dirac equation are:
\begin{eqnarray}
\tilde{\psi}_2&=&k\cos(kx)-\pi\cot(\pi x)\sin(kx),\\
  \tilde{\psi}_1&=&[E-E_1(\bar{E}_1)]\sin(kx),\\
  \tilde{\varphi}_2&=&k\cos(kx-k)-\pi\cot(\pi x)\sin(kx-k),\\
  \tilde{\varphi}_1&=&[E-E_1(\bar{E}_1)]\sin(kx-k).
\end{eqnarray}
Again we have \begin{eqnarray}
              \tilde{\psi}(a)&=& \tilde{\psi}(b)=0, \\
              \tilde{\varphi}(a)&=&\tilde{\varphi}(b)=0
              \end{eqnarray}that makes right part of eq. \eqref{38} to be equal zero that leads
 to evident consequences, similar to those, made in case (i).\\

{\bf Case (iii)}.  The positive branch of the eigenspectrum is the following
\begin{eqnarray}
E_n&=&\sqrt{m^2+k_n^2},\qquad k_n=\frac{\pi}{2}, \qquad n=0,1,...
\end{eqnarray}
and negative one
\begin{eqnarray}
\bar{E}_n&=&-\sqrt{m^2+k_n^2},\qquad k_n=\frac{\pi}{2}, \qquad n=0,1,...
\end{eqnarray}
Eigenfunctions are
\begin{eqnarray}
\phi_1^{(n)}&=&\sin(k_nx),\qquad \phi_2^{(n)}=-\frac{k_n\cos(k_nx)}{E_n+m}, \\
 \bar{\phi}_1^{(n)}&=&\sin(k_nx),\qquad \bar{\phi}_2^{(n)}=-\frac{k_n\cos(k_nx)}{E_n+m}.
\end{eqnarray}
Solutions of  $$h_0\psi=E\psi$$ are chosen as following way
\begin{eqnarray}
\psi_1&=&\sin(kx), \qquad \psi_2=-\frac{k\cos(kx)}{E+m}, \qquad k=\sqrt{E^2-m^2}, \\
\varphi_2(x)&=&\sin(kx-k), \qquad \varphi_1(x)=\frac{k\cos(kx-k)}{E-m}.
\end{eqnarray}
And the full trace of Green function of the transformed Dirac equation is
\begin{eqnarray}
 \label{77} tr\int_0^1 G(x,x)dx&=&-\frac{E}{k}\tan(k).\end{eqnarray}
Taking into account the relation (see Appendix B)
 \begin{eqnarray}\label{2fromB}\tan(k)=\sum_{n=0}^\infty\frac{2k}{(n+1/2)^2\pi^2-k^2}, \end{eqnarray}
we get \begin{eqnarray}
       -\frac{E}{k}\tan(k)&=&\sum_{n=0}^\infty \frac{2E}{E^2-E_n^2}=\sum_{n=0}^\infty\left(\frac{1}{E-E_n}+\frac{1}{E-\bar{E}_n}\right).
       \end{eqnarray}
Again the result \eqref{77} is in agreement with spectral representation result.

Constructing transformation matrix $u$ from eigenfunctions  $\psi^{(1)}(x)$ and $\bar{\psi}^{(1)}(x)$
with the help of simple algebra one can get
\begin{eqnarray}
  u_xu^{-1}&=&\left(
                \begin{array}{cc}
                  \pi\cot(\pi x/2)/2 & 0 \\
                  0 & -\pi\tan(\pi x/2)/2 \\
                \end{array}
              \right),\end{eqnarray}
  \begin{eqnarray}\tilde{\psi}_1(x)=\psi'_1(x)-\frac{\pi}{2}\cot\left(\frac{\pi}{2}x\right)\psi_1=k\cos(kx)-\frac{\pi}{2}\cot\left(\frac{\pi}{2}x\right)\sin(kx),
 \end{eqnarray} \begin{eqnarray}\tilde{\psi}_2(x)=\psi'_2(x)+\frac{\pi}{2}\tan\left(\frac{\pi}{2}x\right)\psi_2=k\cos(kx-k)+\frac{\pi}{2}\tan\left(\frac{\pi}{2}x\right)\sin(kx-k).
\end{eqnarray}
Again we have
\begin{eqnarray}
\tilde{\psi}_1(0)&=&\tilde{\psi}_2(0)=0\longrightarrow\tilde{\psi}(0)=0,\\
\tilde{\varphi}_1(1)&=&\tilde{\varphi}_2(1)=0\longrightarrow\tilde{\varphi}(1)=0,
\end{eqnarray}
from which and Eq. \eqref{38} it follows that transformed eigenfunctions $L\phi^{(n)}$, $L\bar{\phi}^{(n)}$
form complete set.

 The case (iv) is similar to the case (iii). So we omit it's detail discussion.

\section{Discussion}

We have established the relation \eqref{37} that connect the difference of the full traces
for the Green functions of the initial and the Darboux transformed Dirac problems with energies $E_1,$ $E_2$
of initial states, whose wave functions are used for the construction of the transformation matrix
and boundary values of solutions of the initial and the transformed Dirac equations. These
 relation are used to check the completeness of set of wave functions, obtained by the Darboux transformation
 of eigenfunctions of the initial Hamiltonian for some typical boundary problem.
\section*{Acknowledgments}
The author is grateful to the Joint Institute for Nuclear Research
(Dubna, Moscow region, Russia)  for hospitality during this work. I also thank Prof.
B. G. Bagrov for useful discussions. This work
was supported in part by the ``Dynasty'' Fund and Moscow
International Center of Fundamental Physics.

\section*{Appendix A}
Here we will establish the relation between Wronskians of initial and Darboux transformation problem.

Using the identity $\gamma^T\gamma=1$, we represent $\widetilde{W}$ in the form:
\begin{eqnarray}\widetilde{W}&=&\tilde{\psi}^T\gamma\tilde{\varphi}=tr[\gamma\tilde{\varphi}\tilde{\psi}^T]\equiv tr[\gamma\tilde{\varphi}\tilde{\psi}^T
\gamma^T\gamma]=tr[(\gamma\tilde{\varphi})(\gamma\tilde{\psi})^T\gamma].\end{eqnarray}

Now by usage of the definition
\begin{eqnarray}
\tilde{\psi}&=&\psi'-u_xu^{-1}\psi, \\
  \tilde{\varphi}&=&\varphi'-u_xu^{-1}\varphi
\end{eqnarray}
and the Dirac equation
\begin{eqnarray}
\gamma\psi'&=&(E-V)\psi, \qquad \gamma\phi'=(E-V)\phi,
 \end{eqnarray}
 $$ \gamma u_x=u\lambda-Vu$$
we can present $\tilde{\psi}$
and $\tilde{\varphi}$ in the form
\begin{eqnarray}
\tilde{\varphi}&=&(E-\Omega)\varphi,\qquad \tilde{\varphi}=(E-\Omega)\varphi,\qquad \Omega=u\lambda u^{-1}.
\end{eqnarray}
Then we have
\begin{eqnarray}
\widetilde{W}&=&tr[\gamma(E-\Omega)\varphi\psi^T(E-\Omega^T)]\nonumber\\
&=&tr[E^2\gamma\varphi\psi^T-E(\gamma\Omega+\Omega^T\gamma)\varphi\psi^T+\Omega\gamma\Omega\varphi\psi^T].
\end{eqnarray}
 It is easy to verify that for arbitrary $(2\times 2)$ matrix $\Omega$
and $\gamma=\left(
\begin{array}{cc}
 0 & 1 \\
 -1 & 0 \\
\end{array}
\right)$
\begin{eqnarray}
\gamma\Omega+\Omega^T\gamma&=&\gamma tr\Omega
\end{eqnarray}
and
\begin{eqnarray}
\Omega^T\gamma\Omega=\gamma\det\Omega.
\end{eqnarray}
Thus,
\begin{eqnarray}
\widetilde{W}&=&(tr \gamma\varphi\psi^T)(E^2-E tr\Omega+\det\Omega)=W(E^2-E tr\Omega+\det\Omega).
\end{eqnarray}
Further it is easy to check  for arbitrary nonsingular $2\times 2$ matrix $u$ and $$\Lambda=\left(
\begin{array}{cc}
 \lambda_1 & 0 \\
 0 & \lambda_2 \\
 \end{array}
 \right),$$ that
$$tr\Omega=tr u\Lambda u^{-1}=\lambda_1+\lambda_2,$$
$$\det\Omega=\det(u\lambda u^{-1})=\lambda_1\lambda_2.$$

Thus, we obtain the expression related the Wronskians of the transformed and the initial
problems
\begin{eqnarray}
\widetilde{W}&=&(E-\lambda_1)(E-\lambda_2)W,
\end{eqnarray}
where $\lambda_1=E_{n_1}$, $\lambda_2=E_{n_2}$.
\section*{Appendix B}
Here we will prove the relations \eqref{1fromB}, \eqref{2fromB} used in the text  to check the correspondence
of two results for full trace of the Green function of the Dirac problem with boundary conditions.

Let us consider \begin{eqnarray}
                H_1(x)&=&\Gamma(1+x)\Gamma(1-x)\equiv\frac{\pi x}{\sin(\pi x)},
                \end{eqnarray}
where $\Gamma(z)$ is Eiler $\Gamma$--function.
Then \begin{eqnarray}\frac{d\ln{H_1(x)}}{dx}&=&\psi(1+x)-\psi(1-x)=\frac{1}{x}-\pi\cot(\pi x),\\
&&\psi(z)=\frac{d\ln(\Gamma(z))}{dz}.
\end{eqnarray}
Using the representation
\begin{eqnarray}
\psi(z)&=&-C+\sum_{n=1}^\infty\left(\frac{1}{n}-\frac{1}{n-1+z}\right),
\end{eqnarray}
where $C$ is the Eiler constant, it is easy to get
\begin{eqnarray}
\pi\cot{\pi x}&=&\frac{1}{x}+\sum_{n=1}^\infty\frac{2x}{x^2-n^2}.
\end{eqnarray}
By substitution $x=k/\pi$ in the last relation, we have
\begin{eqnarray}
\cot(k)&=&\frac{1}{k}+\sum_{n=1}^\infty\frac{2k}{k^2-\pi^2n^2}.
\end{eqnarray}
Next we consider
\begin{eqnarray}
H_2(x)&=&\Gamma(1/2+x)\Gamma(1/2-x)=\frac{\pi}{\cos(\pi x)},\\
  \frac{\ln{H_2}}{dx}&=&\psi(1/2+x)-\psi(1/2-x)=\pi\tan(\pi x).
\end{eqnarray}
Since \begin{eqnarray}\psi(1/2+x)-\psi(1/2-x)&=&\left[-C+\sum_{n=0}^\infty\left(\frac{1}{n+1}-\frac{1}{n+1/2+x}\right)\right]\nonumber\\
-\left[-C+\sum_{n=0}^\infty\left(\frac{1}{n+1}-\frac{1}{n+1/2-x}\right)\right]&=&\sum_{n=0}^\infty\frac{2x}{(n+1/2)^2-x^2},
\end{eqnarray}
by  substitution $x=k/\pi$ we get
\begin{eqnarray}
\tan(k)&=&\sum_{n=0}^\infty\frac{2k}{\pi^2(n+1/2)^2-k^2}.
\end{eqnarray}


\begin{thebibliography}{0}

\bibitem{Witten} E. Witten,  {\it Nucl. Phys. B}, {\bf 188}, 513
(1981);
\\E. Witten, {\it J. Diff. Geometry.}, {\bf 17}, 661 (1982);\\
 H. Nicolai, {\it J. Phys. A: Math. Gen.}, {\bf 9}, 1497 (1976).


\bibitem{2} G. Junker, {\it Supersymmetric Methods in Quantum and Statstical Physics} (Berlin: Springer 1996);\\
F. Cooper, A. Khare and U. Sukhatme, {\it Phys. Rep.}, {\bf 25} 268 (1995).

\bibitem{3} E. Schr\"o\-din\-ger, {\it Proc. R. Irish Acad.}, {\bf 46} A 9 (1940);\\
E. Schr\"o\-din\-ger, {\it Proc. R. Irish Acad.}, {\bf 46} A 183 (1941);\\
L. Infeld and T.E. Hull, {\it Rev: Mod. Phys.}, {\bf 23} 21 (1951).

\bibitem{4}
G. Darboux, {\it Compt. Le\c{c}ons sur la
th\'{e}orie g\'{e}n\'{e}rale des surfaces et les application
g\'{e}om\'{e}triques du calcul infinit\'{e}simate.}, Paris:
Guatier--Villar et Fils, 522 (1889);\\
M. Crum, {\it J. Math.}, {\bf 6} 121 (1955).


\bibitem{annphys2003v305p151} L. M. Nieto, A. A. Pecheritsin  and
 B. F. Samsonov,  {\it Ann. Phys.}, {\bf 305} 151 (2003);\\
  N. Debergh, A. A. Percheritsin, B. F. Samsonov et
 al.,{\it J. Phys.}, A {\bf 35} 3279 (2002).

\bibitem{Eurjphys}  B. F. Samsonov, A. A. Pecheritsin, E. O. Pozdeeva
et al., {\it Eur. J. Phys.}, {\bf 24} 435 (2003);\\ V. G. Bagrov,   A. A. Pecheritsin, E. O. Pozdeeva
et al., {\it Commun.  Nonlin. Scien.}, {\bf 9} 13 (2004);\\
 E. O. Pozdeeva,  {\it J.  Surf. Invest.}, {\bf 3} 66 (2007).

\bibitem{preprint} E. O. Pozdeeva, {\it Connection between the Green functions for
 the supersymmetric pair  of Dirac Hamiltonians}, arXiv:hep-th/0709.1480.

 \bibitem{Sukumar} C. V. Sukumar,
{\it J. Phys.}, A {\bf 37} 10287 (2004).
\end{thebibliography}
\end{document}